# Interference in Electron-Molecule Elastic Scattering:
## *s*-, *p*- and *d*-spherical waves

A. S. Baltenkov[1], S. T. Manson[2], and A. Z. Msezane[3]

[1] Arifov Institute of Ion-Plasma and Laser Technologies,
100125, Tashkent, Uzbekistan
[2] Department of Physics and Astronomy,
Georgia State University, Atlanta, Georgia 30303, USA
[3] Center for Theoretical Studies of Physical Systems,
Clark Atlanta University, Atlanta, Georgia 30314, USA

**Abstract**
General formulas describing the multiple scattering of electron by polyatomic molecules have been derived within the framework of the model of non-overlapping atomic potentials. These formulas are applied to different carbon molecules, both for fixed-in-space and randomly oriented molecules. The molecular continuum wave function is represented as a plane wave plus a linear combination of the Green's functions for free motion and the derivatives of these functions. Far from the target the electron spherical waves interfere as in the case of the Young's double slits experiment. This interference manifests itself as diffraction oscillations in the differential and total cross sections for elastic electron scattering. The amplitude of electron scattering by a molecule is defined by the phase shifts for each of the atoms forming the target and its geometry. The numerical calculation of the scattering amplitude in closed form (rather than in the form of S-matrix expansion) is reduced to solving a system of algebraic equations. The number of atoms in a molecule and the atomic phase shifts for the orbital angular momenta included in the calculation define the number of equations.

PACS number: 34.80.Bm

## 1. Introduction

A physical picture of a particle wave scattering by a system of spherical scatterers is based on the Huygens-Fresnel principle, according to which in the scattering process the target centers become the sources of the spherically scattered waves and beyond the target there is a system of spherical waves diverging from each of the spatially separated centers. Far from the target, the particle wave function corresponding to this physical picture of scattering has the form [1, 2]

$$\psi_{\mathbf{k}}^{+}(\mathbf{r}) \approx e^{i\mathbf{k}\cdot\mathbf{r}} + \sum_{j=1}^{N} A_j \frac{e^{ik|\mathbf{r}-\mathbf{R}_j|}}{|\mathbf{r}-\mathbf{R}_j|}. \tag{1}$$

Here **k** is the particle wave vector; the vectors $\mathbf{R}_j$ are the positions of the target centers in an arbitrary coordinate system.

As pointed out some time ago [3], the wave function for electron scattering by a molecule is described similarly to Eq. (1). Specifically, "Interference (or diffraction) phenomena should in fact occur when electrons are released within a multi-center molecular field. From Huygens' point of view one may, for example, regard the two atoms of $N_2$ or $O_2$ as essentially independent absorbers of light which constitute, in turn, separate sources of photoelectrons. Superposition of the emissions from these two sources produces an interference pattern whose properties should depend periodically on the ratio of the inter-nuclear distance to the photoelectron wavelength". In addition, electron scattering by a diatomic molecule is discussed in the context of Young's double-slit experiment [4]. It is of interest to analyze how the computational methods for continuum molecular wave functions describing the elastic electron scattering by a molecule correspond to the ideas in [3] and [4]. Furthermore, in the review [5] it is stated, "In a general calculation of molecular wave functions, the molecular orbitals are written in terms of LCAO (linear combination of atomic orbitals) centered at each atom of the molecule. This kind of multi-centered target wave function is very inconvenient to use in any e-molecule scattering equation. It is, therefore, more convenient if all the quantities, such as the bound and the continuum orbitals, interaction potentials, *etc*, are expanded around a single center, normally the center-of-mass (COM) of the system. Then, the scattering equations for the continuum electron function *will be similar to those for electron-atom scattering* and will become computationally more tractable". In the description of the continuum multiple scattering $CMSX_\alpha$ method [6-8] we find [8], "In the electron scattering case, *the continuum electron is treated as moving in a spherically averaged molecular field* which is divided into an inner region, chosen from the electron densities of the isolated component atoms, and an outer region encompassing the whole molecule". The asymptotic behavior of the continuum wave function in this scattering picture is described by the following expression, Eq.(71) in [5] as

$$\psi_{\mathbf{k}}^+(\mathbf{r} \to \infty) = e^{i\mathbf{k}\cdot\mathbf{r}} + F(\mathbf{k},\mathbf{k}')\frac{e^{ikr}}{r}. \qquad (2)$$

where **r** is the electron radius vector in the coordinate system with origin at the COM of the target, **k** and $\mathbf{k}' = k\mathbf{r}/r$ are the initial and final directions of the continuum electron and $F(\mathbf{k},\mathbf{k}')$ is the elastic scattering amplitude.

Much more sophisticated calculations of the molecular continuum [9], modified versions of the multiple scattering method in which the atomic spheres overlap [10-13], *ab initio* methods like the Hartree–Fock approximation with subsequent inclusion of many-electron correlations [14, 15] or the multi-channel Schwinger configuration interaction method [16, 17], *etc.* – all these impose the condition that the electron continuum wave function "must have the asymptotic form of an incident *plane wave* plus *outgoing spherical waves*" [8] that correspond to the spherically symmetric target and is described by Eq. (2).

The single center expansion of the continuum orbital is described in [5] as analogous to the expansion of the bound orbitals. However, there are inherent problems with this methodology. Originally, the single center expansion was used in molecular physics to calculate the bound state eigenvalues [18]. In the calculation of the bound state wave functions the asymptotic form and normalization and is evident. The situation with the continuum wave functions is quite different. Specifically, before the expansion relative to the COM is carried out, we must choose the asymptotic behavior of the continuum wave function, i.e., the form of equations (1) or (2). Evidently, the difference between the pictures of electron wave scattering, equations (1) and (2) has significant consequences. In the first case, we deal with the multi-center picture of scattering while in the second case, the phenomenon of electron diffraction by molecules in its classical description as interference of the spherical waves generated by the spatially separated centers is impossible because, beyond the target, there is only a single spherical wave, and this is clearly unphysical. Thus, for computational convenience the real potential field of the multi-center target is replaced by a spherically averaged molecular field; the wave diffraction by a system of centers is replaced by the diffraction on a single molecular sphere.

Another approach to the electron-molecule scattering problem – the independent-atom model [19-22] - reduces this problem to the electron-atom scattering. In this approach the molecular elastic scattering amplitude is represented as a sum of those for the component atoms with a suitable phase factor due to their different locations in the molecule. In this model the differential cross section of elastic electron scattering averaged over all the orientations of the molecular axes relative to electron beam has the form

$$\frac{d\sigma}{d\Omega} = \sum_{i=1}^N \frac{d\sigma_i}{d\Omega} + \sum_{i \ne j=1}^N F_i^*(\vartheta) F_j(\vartheta) \frac{\sin(sr_{ij})}{sr_{ij}}, \qquad (3)$$

where $d\sigma/d\Omega$ and $d\sigma_i/d\Omega$ are the molecular and atomic differential cross sections, respectively; $F_i(\vartheta)$ is the atomic scattering amplitude; $r_{ij}$ are the distances between *i*-th and *j*-th atoms in the target; $N$ is the total number of atoms present in the molecule. The function $s = 2k\sin(\vartheta/2)$ in (3) is the magnitude of the momentum transfer in the collisions. This approach is widely used for electron-molecule collisions with $F_i(\vartheta)$ usually obtained from atomic scattering phase shifts. While this treatment accounts for the interference of the scattering by individual atoms, it ignores multiple scattering, which is the objective of present work.

The method developed in [23] for calculating the continuum wave function differs fundamentally from the above-mentioned traditional consideration of the electron elastic scattering process in molecular physics. This method makes it possible to analyze diffraction phenomena in electron scattering by a *system of scatterers* fixed in space rather than diffraction by an *isolated spherically averaged molecular field*. According to [23], in the problem of electron scattering by a cluster of atomic potentials, the wave function corresponding to Eq. (1) can be represented as a plane wave and a linear combination of the Green's functions of the free motion and the derivatives of these functions. The boundary conditions imposed on these functions at the centers of the atomic potentials result in a system of inhomogeneous algebraic equations. Their solution defines the coefficients of the linear combination and the amplitudes of electron scattering by a target. Thus, in [23] it was shown that if one knows the scattering phases for each of the short-range atomic potentials forming the target, and the target geometry, then the amplitude of electron elastic scattering could be written in closed form rather than in the form of partial-wave expansion.

Note that a similar method was used in nuclear physics by Brueckner [2] when considering multiple scattering by two-center targets. There the wave function of meson scattering by deuterons was constructed as a plane wave and a combination of the Green's functions (*s*-waves) and their gradients (*p*-waves). Indeed, there it was shown that the consideration of the *p*-waves leads to significant corrections in the scattering cross section as compared to the result of only *s*-wave scattering calculation. In [23], a systematic method was developed to build the continuum wave function with *s, p, d… etc,* orbital angular momenta of outgoing waves and applied to the case of two-atom molecules created by identical atoms, homonuclear diatomics.

In the present paper we generalize the methodology and the equations of [23] and [24] to the calculation of cross sections of elastic scattering of slow electrons by any molecule using the model of non-overlapping atomic potentials. In Section 2 the formulas [23, 24] derived for two-atom molecules with identical atoms are generalized to describe electron scattering by targets with an arbitrary number of atoms. These general formulas are applied in Section 3 to CH molecules fixed-in-space and randomly oriented relative to an electron beam. In Sections 4 and 5 the derived formulas are used for carbon molecules with toroidal symmetry (5-ring and 6-ring carbons) as well as for molecules with cylindrical symmetry (linear chains of carbon atoms). In Section 6 the analogy between electron-molecule scattering and the Young's double slits experiment is briefly discussed. Section 7 presents the Conclusions. Appendix A is the final of the paper.

## 2. General formulas

We represent a molecule as a cluster of $N$ non-overlapping atomic spheres with the centers located at the points $\mathbf{R}_i$. We assume that the molecule neither vibrates nor rotates during its collision with the electron. Consider the electron continuum wave function in the field of a single atom of the target. *Beyond the atomic sphere* centered at $\mathbf{r} = \mathbf{R}_i$ the electron wave function consists of a combination of the usual spherical Bessel functions [25, 26] (atomic units are used throughout the paper)

$$\psi_{\mathbf{k}}^{+}(\mathbf{r}) = \sum_{l,m} C_{lm}^{(i)}[j_l(k\rho_i)\cot\delta_l^{(i)} - n_l(k\rho_i)]Y_{lm}(\mathbf{\rho}_i), \tag{4}$$

where $\mathbf{\rho}_i = \mathbf{r} - \mathbf{R}_i$. The scattering phase of the *l*-th partial wave $\delta_l^{(i)}$ in equation (4) is a function of $k$ and is assumed to be known for each scattering center. Here and throughout the paper $Y_{lm}(\mathbf{r}) \equiv Y_{lm}(\vartheta,\varphi)$ where $\vartheta$ and $\varphi$ are the spherical angles of the vector $\mathbf{r}$. We define the wave function, equation (4), for *all r* and find the boundary conditions that should be imposed on the wave function at the point *r*=0 so that beyond the atomic sphere it would be described by formula (4). In the limit $\mathbf{\rho}_i \to 0$, equation (4) leads to the boundary condition for the wave function at $\mathbf{r} = \mathbf{R}_i$

$$\psi_{\mathbf{k}}^{+}(\mathbf{r})_{\rho_i \to 0} \approx \sum_{l,m} C_{lm}^{(i)}[A_{kl}\rho_i^l \cot\delta_l^{(i)} - B_{kl}\rho_i^{-l-1}]Y_{lm}(\mathbf{\rho}_i), \tag{5}$$

where $A_{kl} = k^l/(2l+1)!!$ and $B_{kl} = -(2l-1)!!/k^{l+1}$. The coefficients $C_{lm}^{(i)}$ depend on the behavior of the wave function $\psi_{\mathbf{k}}^{+}$ at large distance *r*. We can express the coefficients $C_{lm}^{(i)}$ in terms of $\psi_{\mathbf{k}}^{+}(\mathbf{r})$ as

$$C_{lm}^{(i)} = \tan\delta_l^{(i)} \frac{(2l+1)!!}{k^l(2l+1)!}\left[\left(\frac{d}{d\rho_i}\right)^{2l+1} \rho_i^{l+1}\int \psi_{\mathbf{k}}^{+}(\mathbf{r})Y_{lm}^{*}(\mathbf{\rho}_i)d\Omega_{\rho_i}\right]_{\rho_i \to 0}. \tag{6}$$

Here $d\Omega_{\rho_i}$ is the element of the solid angle of the vector $\mathbf{\rho}_i$. Thus defined, the coefficients $C_{lm}^{(i)}$ do not depend on the normalization of the wave function $\psi_{\mathbf{k}}^{+}$.

Since the interaction of the scattered electron with each of the target atoms has a short-range character in the present approximation, and does not depend on other atoms, then the wave function near each of the atoms has the same form as in the absence of the neighboring atom. Therefore, the wave function of the electron scattered by a cluster of the atomic potentials should obey the boundary conditions (5) and (6). Consistent with Eq. (1), we write the wave function of electron scattering by a target as

$$\psi_{\mathbf{k}}^{+}(\mathbf{r}) = e^{i\mathbf{k}\cdot\mathbf{r}} + \sum_{j=1}^{N}\sum_{\lambda,\mu} D_{\lambda\mu}^{(j)} P_{k\lambda}(\rho_j)Y_{\lambda\mu}(\mathbf{\rho}_j). \tag{7}$$

In equation (7), the wave function representing electron scattering has the form of a plane wave plus $N$ spherical waves, $P_{k\lambda}(|\mathbf{\rho}_j|)Y_{\lambda\mu}(\mathbf{\rho}_j)$, emerging from the $N$ scattering centers; the amplitudes of these spherical waves are determined by the unknown coefficients $D_{\lambda\mu}^{(j)}$. The radial parts of the spherical waves in Eq. (7) have the form [26]

$$P_{k\lambda}(\rho_j) = ik[j_\lambda(k\rho_j) + in_\lambda(k\rho_j)] = ikh_\lambda(k\rho_j) = (-1)^\lambda \frac{\rho_j^\lambda}{k^\lambda}\left(\frac{1}{\rho_j}\frac{d}{d\rho_j}\right)^\lambda \frac{e^{ik\rho_j}}{\rho_j} \tag{8}$$

where $h_\lambda(x)$ is the Hankel function [25]. For $\mathbf{\rho}_j \to 0$ the radial parts of the wave functions (8) are described by

$$P_{k\lambda}(\rho_j) = ik\left[A_{k\lambda}\rho_j^\lambda + iB_{k\lambda}\rho_j^{-\lambda-1}\right] \approx -B_{k\lambda}k\rho_j^{-\lambda-1}. \tag{9}$$

The number of the spherical $\lambda\mu$-waves in the wave function (7) is, in principle, infinite. However, the partial wave expansion for slow particles converges rapidly so that the summation needs to include orbital only angular momentum $0 \leq \lambda \leq l_{max}$, where $l_{max}$ is the highest partial wave considered. Each value of the orbital angular momentum $\lambda$ corresponds to $2\lambda+1$ spherical waves, each with the different magnetic quantum number $\mu$. Therefore, the number of $\lambda\mu$-waves emitted by each of the centers is $(l_{max}+1)^2$. Hence, the number of the coefficients $D^{(j)}_{\lambda\mu}$ defining the amplitudes of the spherical waves in the wave function (7) is $N(l_{max}+1)^2$.

For the wave function (4) to satisfy the boundary condition (5), it is necessary, in equations (4) and (7), to equate the terms approaching infinity at the point $\mathbf{r} \to \mathbf{R}_i$. This requires that the coefficients $C^{(i)}_{lm}$ and $D^{(i)}_{lm}$ be connected through $C^i_{lm} = kD^{(i)}_{lm}$. Hence, the amplitudes of the spherical waves $D^{(i)}_{lm}$ in equation (7) are described by

$$D^{(i)}_{lm} = \tan\delta^{(i)}_l \frac{(2l+1)!!}{k^{l+1}(2l+1)!} \left[ \left(\frac{d}{d\rho_i}\right)^{2l+1} \rho_i^{l+1} \int \psi^+_{\mathbf{k}}(\mathbf{r}) Y^*_{lm}(\boldsymbol{\rho}_i) d\Omega_{\rho_i} \right]_{\rho_i \to 0}. \tag{10}$$

For the calculation of these coefficients, we rewrite the wave function, equation (7), as

$$\psi^+_{\mathbf{k}}(\mathbf{r}) = e^{i\mathbf{k}\cdot\mathbf{r}} + D^{(i)}_{\lambda\mu} P_{k\lambda}(\rho_i) Y_{\lambda\mu}(\boldsymbol{\rho}_i) + \sum_{\substack{j=1,\\j\neq i}}^{N}\sum_{\lambda,\mu} D^{(j)}_{\lambda\mu} \hat{B}^{(j)}_{\lambda\mu} G^+_k(\mathbf{r},\mathbf{R}_j). \tag{11}$$

The spherical waves with arguments $\boldsymbol{\rho}_j \neq \boldsymbol{\rho}_i$ in equation (11) are the result of applying the operators $\hat{B}^{(j)}_{lm}$ on the Green's function for free motion $G^+_k(\mathbf{r},\mathbf{R}_j)$ (see below Appendix A)

$$\hat{B}_{lm} G^+_k(\mathbf{r},\mathbf{R}_j) = P_{kl}(\rho_j) Y_{lm}(\boldsymbol{\rho}_j). \tag{12}$$

Substituting the wave function (11) into equation (10), and integrating over the solid angles of the vectors $\boldsymbol{\rho}_i$, we obtain the following two limits for $D^{(i)}_{lm}$,

$$L_1 = \left[ \left(\frac{d}{d\rho_i}\right)^{2l+1} \rho_i^{l+1} \int e^{i\mathbf{k}\cdot\mathbf{r}} Y^*_{lm}(\boldsymbol{\rho}_i) d\Omega_{\rho_i} \right]_{\boldsymbol{\rho}_i \to 0} = 4\pi i^l \frac{k^l(2l+1)!}{(2l+1)!!} e^{i\mathbf{k}\cdot\mathbf{R}_i} Y^*_{lm}(\mathbf{k}), \tag{13}$$

$$L_2 = \left[ \left(\frac{d}{d\rho_i}\right)^{2l+1} \rho_i^{l+1} D^{(i)}_{lm} P_{kl}(\rho_i) \right]_{\rho_i \to 0} = \frac{k^l(2l+1)!}{(2l+1)!!} D^{(i)}_{lm} ik. \tag{14}$$

For the third limit

$$L_3 = \left[ \left(\frac{d}{d\rho_i}\right)^{2l+1} \rho_i^{l+1} \sum_{\substack{j=1,\\j\neq i}}^{N}\sum_{\lambda,\mu} \int D^{(j)}_{\lambda\mu} \hat{B}^{(j)}_{\lambda\mu} G^+_k(\mathbf{r},\mathbf{R}_j) Y^*_{lm}(\boldsymbol{\rho}_i) d\Omega_{\rho_i} \right]_{\rho_i \to 0}, \tag{15}$$

where the arguments of the Green's function were replaced, $G^+_k(\mathbf{r},\mathbf{R}_j) = G^+_k(\boldsymbol{\rho}_i,\mathbf{R}_{ji})$, with $\mathbf{R}_{ji} = \mathbf{R}_j - \mathbf{R}_i$. With this the integration can be performed over the solid angles of the vector $\boldsymbol{\rho}_i$ taking into account that for $\rho_i < R_{ji}$ the Green's function can be written as [27]

$$G^+_k(\boldsymbol{\rho}_i,\mathbf{R}_{ji}) = \frac{e^{ik|\boldsymbol{\rho}_i-\mathbf{R}_{ji}|}}{2\pi|\boldsymbol{\rho}_i-\mathbf{R}_{ji}|} = 2ik\sum_{l_1} j_{l_1}(k\rho_i) h_{l_1}(kR_{ji}) \sum_{m_1=-l_1}^{l_1} Y_{l_1 m_1}(\boldsymbol{\rho}_i) Y^*_{l_1 m_1}(\mathbf{R}_{ji}). \tag{16}$$

Consequently, we have, for the third limit,

$$L_3 = 2ik \left[ \frac{(2l+1)!k^l}{(2l+1)!!} \sum_{\substack{j=1, \\ j \neq i}}^{N} \sum_{\lambda,\mu} D_{\lambda\mu}^{(j)} \hat{B}_{\lambda\mu}^{(j)} [h_l(kR_{ji}) Y_{lm}^*(\mathbf{R}_{ji})] \right]. \tag{17}$$

Collecting together equations (10), (13), (14) and (17), we obtain a system of linear algebraic equations for the $N(l_{max}+1)^2$ coefficients, $D_{lm}^{(i)}$,

$$D_{lm}^{(i)} - 2ik f_l^{(i)} \sum_{\substack{j=1, \\ j \neq i}}^{N} \sum_{\lambda=0}^{l_{max}} \sum_{\mu=-\lambda}^{\lambda} D_{\lambda\mu}^{(j)} \hat{B}_{\lambda\mu}^{(j)} [h_l(kR_{ji}) Y_{lm}^*(\mathbf{R}_{ji})] = 4\pi i^l f_l^{(i)} e^{i\mathbf{k}\cdot\mathbf{R}_i} Y_{lm}^*(\mathbf{k}). \tag{18}$$

Here the differential operators $\hat{B}_{\lambda\mu}^{(j)}$ act on the functions inside the square brackets and the function $f_l^{(i)} = \exp[i\delta_l^{(i)}]\sin\delta_l^{(i)}/k$ is the partial amplitude for elastic scattering by the $i$-th center. The number of linear equations in (18) is $N(l_{max}+1)^2$. In the case of two identical atomic spheres, with the centers at the points $\mathbf{R}_1 = -\mathbf{R}/2$ and $\mathbf{R}_2 = \mathbf{R}/2$, equations (18) reduce to the equations (24) and (25) of Ref. [23].

Calculated using equations (18), the coefficients $D_{lm}^{(i)}$ define the molecular continuum wave function (11). At large distances from target the continuum wave function has the form

$$\psi_{\mathbf{k}}^+(\mathbf{r} \to \infty) = e^{i\mathbf{k}\cdot\mathbf{r}} + \sum_{j=1}^{N} \sum_{\lambda,\mu} D_{\lambda\mu}^{(j)} \hat{B}_{\lambda\mu}^{(j)} \frac{e^{ik|\mathbf{r}-\mathbf{R}_j|}}{2\pi|\mathbf{r}-\mathbf{R}_j|} = e^{i\mathbf{k}\cdot\mathbf{r}} + \frac{e^{ikr}}{r} \frac{1}{2\pi} \sum_{j=1}^{N} \sum_{\lambda,\mu} D_{\lambda\mu}^{(j)} \hat{B}_{\lambda\mu}^{(j)} e^{-i\mathbf{k}'\cdot\mathbf{R}_j}. \tag{19}$$

From equation (19), the amplitude for elastic electron scattering by a target becomes

$$F(\mathbf{k},\mathbf{k}',\{\mathbf{R}_j\}) = \frac{1}{2\pi} \sum_{j=1}^{N} \sum_{\lambda,\mu} D_{\lambda\mu}^{(j)} \hat{B}_{\lambda\mu}^{(j)} e^{-i\mathbf{k}'\cdot\mathbf{R}_j}. \tag{20}$$

The differential elastic scattering cross section $d\sigma/d\Omega_{k'} = |F(\mathbf{k},\mathbf{k}',\{\mathbf{R}_j\})|^2$ is a function of the relative orientation of the vectors **k** and **k**' with respect to the set of molecular axes, $\{\mathbf{R}_j\}$. The total cross section is defined by the optical theorem [26] according to which

$$\sigma_{tot}(\mathbf{k},\{\mathbf{R}_j\}) = \int \frac{d\sigma}{d\Omega_{k'}} d\Omega_{k'} = \frac{4\pi}{k} \operatorname{Im} F(\mathbf{k}=\mathbf{k}',\{\mathbf{R}_j\}) = \frac{2}{k} \operatorname{Im} \sum_{j=1}^{N} \sum_{\lambda,\mu} D_{\lambda\mu}^{(j)} \hat{B}_{\lambda\mu}^{(j)} e^{-i\mathbf{k}\cdot\mathbf{R}_j}. \tag{21}$$

The total cross section, $\sigma_{tot}$, depends on the relative orientation of the molecular axis and the incident electron beam momentum **k**.

In the following sections we apply these general formulas to a variety of molecular systems, both homonuclear and heteronuclear. In Sections from 3 to 6 we will restrict ourselves considering only *s*-scattering by each of atoms of the target. The aim of these numerical calculations is to illustrate the possibilities of derived formulas for targets with *N*>2.

## 3. Diatomic molecules

If we apply equation (18) to a heteronuclear diatomic molecule with *s*-wave scattering amplitudes $f_0^{(1)} = \exp[i\delta_0^{(1)}]\sin\delta_0^{(1)}/k$ and $f_0^{(2)} = \exp[i\delta_0^{(2)}]\sin\delta_0^{(2)}/k$, then we will have (for $l_{max} = 0$) [24]

$$F(\mathbf{k},\mathbf{k}',\{\mathbf{R}_i\}) = \frac{1}{a^2 - b_1 b_2} \{b_2 e^{i(\mathbf{k}-\mathbf{k}')\cdot\mathbf{R}_1} + b_1 e^{i(\mathbf{k}-\mathbf{k}')\cdot\mathbf{R}_2} - a[e^{-i(\mathbf{k}'\cdot\mathbf{R}_1 - \mathbf{k}\cdot\mathbf{R}_2)} + e^{-i(\mathbf{k}'\cdot\mathbf{R}_2 - \mathbf{k}\cdot\mathbf{R}_1)}]\}. \tag{22}$$

Here the following designations are used

$$a = \exp(ikR)/R, \quad b_{1,2} = -1/f_0^{(1,2)}, \tag{23}$$

where $\mathbf{R} = \mathbf{R}_1 - \mathbf{R}_2$. Note that the restriction to *s*-wave scattering is relevant to very low energy scattering. The differential elastic scattering cross section $d\sigma/d\Omega_{k'}$ is extremely complicated, even for this case. Therefore, we consider the total cross section, equation (21), described in this case by

$$\sigma_{tot}(\mathbf{k},\mathbf{R}) = \frac{4\pi}{k} \text{Im} \frac{1}{a^2 - b_1 b_2}[b_1 + b_2 - 2a\cos(\mathbf{k}\cdot\mathbf{R})]. \tag{24}$$

The total cross section $\sigma_{tot}(\mathbf{k},\mathbf{R})$ (24) averaged over all the directions of the momentum of the incident electron $\mathbf{k}$ relative to the vector $\mathbf{R}$ is then

$$\bar{\sigma}(k) = \frac{1}{4\pi}\int \sigma_{tot}(\mathbf{k},\mathbf{R}) d\Omega_k = \frac{4\pi}{k} \text{Im}\left[\frac{b_1 + b_2 - 2aj_0(kR)}{a^2 - b_1 b_2}\right], \tag{25}$$

where $j_0(x) = \sin x/x$ is the spherical Bessel function [25]. In the case of identical atoms, $b_1 = b_2$ and equations (24) and (25) then coincide with equations (16) and (18) of Ref. [24].

We now apply equations (24) and (25) to slow elastic electron scattering by CH molecules both fixed and randomly oriented in space, using internuclear distance for the CH molecule $R$=1.1198 Å=2.116 au [28]. In the parameter $b_1$ the function $\delta_0^{(1)}(k) = \delta_0^C(k)$ for a single carbon atom is well-described by $\delta_0^{(1)}(k) \approx 2\pi - 1.912k$ [24]. In the case of a single H atom, for the s-phase shifts $\delta_0^{(2)}(k) = \delta_0^H(k)$ defining the parameter $b_2$, we have two sets of phase shifts: triplet and singlet phase shifts. They are very well described by the expressions: $\delta_0^t(k) = \pi - 1.92564k$ and $\delta_0^s(k) = \pi - 5.72682k + 3.62932k^2$ [29]. With these sets of phases we calculate the cross sections of electron scattering by both CH-triplet and CH-singlet molecules.

The total $\sigma_{tot}(\mathbf{k},\mathbf{R})$, equation (24), and averaged $\bar{\sigma}(k)$, equation (25), cross sections for these molecules as functions of electron momentum $k$ are presented in figure 1 where the polar angle $\vartheta_R$ is defined as the position of the vector $\mathbf{R}$ relative to the z-axis i.e., the coordinate system is oriented so that $\mathbf{k}$ is along the z-axis. In both the panels of figure 1 the curves $\sigma_{tot}(\mathbf{k},\mathbf{R})$ for different angles $\vartheta_R$ merge in the limit $k \to 0$ when the electron wavelength $\lambda = 1/k$ significantly exceeds the molecular size. The waves of such a length are scattered by a target as if it is a point source of the spherical waves. The limiting values of the cross sections $\sigma_{tot}(0,\mathbf{R})$ for CH-singlet and CH-triplet molecules are different, because they are defined by the singlet and triplet lengths of electron scattering by hydrogen atom. The curves corresponding to the averaged cross section $\bar{\sigma}(k)$ almost coincide with the curves $\sigma_{tot}(\mathbf{k},\mathbf{R})$ for $\vartheta_R = 60^o$. The trigonometric functions $\sin kR/kR$ in equations (24) and (25) lead to diffraction maxima when the electron wavelength becomes equal to the distance between the atoms, $kR \approx 1$. They also make these cross sections approach zero for $kR \approx \pi$.

## 4. 6- and 5-rings of carbon atoms

Applying equation (18) to carbon rings, the calculation of the scattering amplitudes for the case of $l_{max} = 0$ is reduced to finding the solutions of the equations

$$D_{00}^{(i)} - f_0^{(i)} \sum_{j=1, j\neq i}^{N} D_{00}^{(j)} \frac{\exp(ikR_{ji})}{R_{ji}} = \sqrt{4\pi} f_0^{(i)} e^{i\mathbf{k}\cdot\mathbf{R}_i}. \tag{26}$$

Explicitly, this system of equations has the form for the 6-ring carbon atoms

$$\begin{Vmatrix} 1 & a_{12} & a_{13} & a_{14} & a_{15} & a_{16} \\ a_{21} & 1 & a_{23} & a_{24} & a_{25} & a_{26} \\ a_{31} & a_{32} & 1 & a_{34} & a_{35} & a_{36} \\ a_{41} & a_{42} & a_{43} & 1 & a_{45} & a_{46} \\ a_{51} & a_{52} & a_{53} & a_{54} & 1 & a_{56} \\ a_{61} & a_{62} & a_{63} & a_{64} & a_{65} & 1 \end{Vmatrix} \cdot \begin{Vmatrix} D_{00}^{(1)} \\ D_{00}^{(2)} \\ D_{00}^{(3)} \\ D_{00}^{(4)} \\ D_{00}^{(5)} \\ D_{00}^{(6)} \end{Vmatrix} = \begin{Vmatrix} S_1 \\ S_2 \\ S_3 \\ S_4 \\ S_5 \\ S_6 \end{Vmatrix}.$$  (27)

Here the coefficients of the unknown functions $D_{00}^{(i)}$ are defined as

$$a_{ji} = -f_0^{(1)} \frac{\exp(ikR_{ji})}{R_{ji}}; \quad S_i = \sqrt{4\pi} f_0^{(1)} e^{i\mathbf{k}\cdot\mathbf{R}_i}.$$  (28)

The amplitude of scattering by a carbon atom, $f_0^{(1)}$ is defined above as $f_0^{(1)} = \exp[i\delta_0^{(1)}]\sin\delta_0^{(1)}/k$. In the spherical coordinate system with the origin at the carbon ring center and the vector **n** as a polar axis (normal to the plane where the rings are located) the coordinates of atomic nuclei of the 6-ring ($\rho_6, \vartheta_R, \varphi_R$) and 5-ring ($\rho_5, \vartheta_R, \varphi_R$) carbons are given in Table 1. The atoms of the 6-ring carbons are located at the circle of the radius $\rho_6$ equal to the distance between the carbon atoms $\rho_6 = R = 1.440$Å $= 2.7211$ au [30]. The atoms of the 5-ring carbons are located at the circle of radius $\rho_5 = R/(2\sin 36^o) = 2.3147$ au, i.e. the distances between the nuclei of the C atoms in both rings are assumed to be equal to $R$.

Table 1. Spherical coordinates of the C-atoms in the 6- and the 5-rings carbons

|       | $\rho_6$, au | $\vartheta_R$, rad | $\varphi_R$, rad | $\rho_5$, au | $\vartheta_R$, rad | $\varphi_R$, rad |
|-------|--------------|--------------------|------------------|--------------|--------------------|------------------|
| $R_1$ | 2.7211       | $\pi/2$            | 0.               | 2.3147       | $\pi/2$            | 0.               |
| $R_2$ |              |                    | $\pi/3$          |              |                    | $2\pi/5$         |
| $R_3$ |              |                    | $2\pi/3$         |              |                    | $4\pi/5$         |
| $R_4$ |              |                    | $\pi$            |              |                    | $6\pi/5$         |
| $R_5$ |              |                    | $4\pi/3$         |              |                    | $8\pi/5$         |
| $R_6$ |              |                    | $5\pi/3$         |              |                    |                  |

The numerical solution of equations (27) with these $R_j$ is reduced to the transformation of the appropriate determinants into a triangular form. The roots of the system of equations are used to calculate the elastic scattering amplitudes, equation (20), and the total cross sections, equation (21). The numerical results for the total cross section $\sigma_{tot}(\mathbf{k},\{\mathbf{R}_j\})$ as a function of electron momentum $k$ for different positions of the rings relative to the incident electron beam are presented in figure 2. The curves in this figure correspond to rotation of the plane with 6- and 5-rings carbons relative to the vector **k**; $\vartheta_n$ is the angle between the vectors **n** and **k** (see insert in the lower panel of figure 2). For that rotation the vectors $\mathbf{R}_1$ and $\mathbf{R}_4$ (of the 6-ring carbons) and the vector $\mathbf{R}_1$ (of the 5-ring carbons) are in the same plane as the vectors **n** and **k**.

The qualitative behavior of the curves in figures 1 and 2 is the same. The more complicated structure of the curves in figure 2 is due to the diffraction oscillation of the cross sections as a result of the interference of the 6 (or 5) spherical *s*-waves. These waves are emitted by centers spatially separated by the different distances $R_{ij}$. In the case of two-center targets (seen in figure 1) we deal with only the parameter $kR$ defining the periodicity of the interference pattern.

## 5. Linear chains of carbon atoms

The elastic scattering cross section by a linear chain of carbon atoms, using $C_6$ as an example, is calculated using equation (26). Taking the unit vector **j** along the molecular axis, the coordinates of the scattering centers are described by $\mathbf{R}_i = \mathbf{j}R(i-1)$, i.e., the interatomic distance in the chain is taken to be the same as in carbon rings and the first atom of the chain is located at the origin of the coordinate system. These solutions are then used for the solution of equation (27), and the total cross sections, equation (21) are obtained as a function of the vectors $\mathbf{R}_i$. The results for $\sigma_{tot}(\mathbf{k},\{\mathbf{R}_j\})$, as a function of the electron momentum $k$ for two different positions of the chain axis **j** relative to the incident electron beam **k** are presented in figure 3. For comparison the total elastic scattering cross section by a 6-ring carbon is also given in figure 3. In these calculations, the mutual orientation of the chain's **j** axis and the 6-ring carbons is given in the insert in figure 3. The diffraction effects in $\sigma_{tot}(\mathbf{k},\{\mathbf{R}_j\})$ manifest themselves more vividly for electron scattering by an atomic chain being parallel to the vector **k** ($\vartheta_j = 0^o$). According to figure 3, in the limit $k \to 0$ the scattering cross sections of the linear molecule $C_6$ and the same molecule coiled up to compact 6-rings, differ by almost a factor of two.

## 6. Electron-molecule elastic scattering and Young's slits experiment

If the axis of the linear atomic chain **j** is perpendicular to an incident electron beam **k** (in our case along the x-axis) then electron scattering by this target very much resembles Young's slit experiment. The difference between these processes is that in electron-molecule scattering the multiple scattering of electron wave (the second term in the left side of equation (18)) plays an important role. It can be neglected when the electron wavelength is small as compared with the distances between the centers. In the Young's slits experiment, on the other hand, there is no multiple scattering of waves by the slits. Omitting the second term in the left side of equation (18), we come to the following equations for the amplitudes of the spherical waves $D_{lm}^{(i)}$ in the electron scattering process in the spirit of the Young's slits experiment

$$D_{lm}^{(i)} = 4\pi i^l f_l^{(i)} e^{i\mathbf{k}\cdot\mathbf{R}_i} Y_{lm}^*(\mathbf{k}) . \tag{29}$$

The differential elastic scattering cross sections for a chain of carbon atoms both including multiple scattering, equation (18), and in the single wave scattering approximation, equation (29), are presented in figure 4 for electron momentum $k$=0.9 au. In the upper panel of figure 4, the results for single scattering approach are given for 3-6 atomic chains; in the lower panel are the same cross sections including the multiple scattering processes. With the increase in the number of atoms in the chain, $N$ (or slits in the diffraction grid), in the diffraction patterns the intensity of the central peak as compared to the secondary peaks rapidly increases, as it should be for Fraunhofer diffraction [31]. The comparison of the giant peak intensities in the upper and lower panel in figure 4 shows that multiple scattering decreases the cross sections $d\sigma/d\Omega_{k'}$ by about a factor of two.

## 7. Conclusions

The consideration of a molecule as a cluster of non-overlapping atomic spheres is a widely used approach in molecular physics despite the fact that this approach is an essential simplification of the real molecular field. This model is usually combined with an additional assumption; beyond the molecular sphere (that encloses the atomic spheres), the molecular field is spherically symmetric. Consequently, far from the molecule, the continuum wave function is the sum of a plane wave plus a single spherical wave emitted from the COM of the target, equation (2). The phase shifts of this wave function are defined from the matching conditions for the solutions of the wave equation on the surfaces of the atomic and molecular spheres. Thus, the problem of electron scattering by a *spherically non-symmetrical* potential is reduced to the conventional S-matrix method of partial waves for a *spherical scatterer*.

The approach employed in [23, 24] is based on the multi-center scattering picture. According to which, in the electron scattering process the atomic spheres become the sources of the scattered waves and far from the target there is a system of spherical waves emerging from each of the spatially separated centers, equation (1). Evidently, the difference between the continuum wave functions that have the differing asymptotic forms of equation (1) and equation (2) has consequences. The first case leads to the multi-center picture of scattering (see Fig. 1a of [32]), while in the second case, the phenomenon of electron diffraction by molecules as interference of spherical waves becomes impossible because beyond the target, there is only a single spherical wave (Fig. 1c of [32]).

In the present paper, elastic scattering of an electron by polyatomic molecules has been evaluated for fixed-in-space and randomly oriented carbon molecules. The general formulas describing the multiple scattering of an electron wave by a target were applied to diatomic molecules and carbon molecules with toroidal and cylindrical symmetry. The wave function of the molecular continuum was represented as a plane wave and a linear combination of the Green's functions of free motion and the derivatives of these functions, equation (11). This construction leads to molecular continuum wave functions that are the asymptotically correct solutions of the wave equation beyond the non-overlapping atomic spheres (see [23] for details). Far from the target, the electron spherical waves interfere as in the case of the Young's double slit experiment; this interference manifests as oscillations in the elastic electron scattering cross sections. The particular features of this diffraction phenomenon are totally defined by the *continuum wave function*. Note that the periodic modulations in the molecular photoionization cross sections that are usually treated as molecular interference [3] have a completely different nature; they are due to the multicenter structure of the *initial state* molecular wave function [32].

Further, we have demonstrated that, if the elastic scattering phase shifts for each of the individual atoms forming a molecular target, and the target geometry are known, then we can write the elastic scattering amplitude *in closed form*, equation (20), and calculate the amplitude by solving a system of algebraic equations, equation (18). The number of these equations rapidly increases with the increase in the number of scattering centers, $N$, and the number of partial waves, $l_{max}$, taken into account in the calculations. This naturally complicates the calculations. Nevertheless, complexity is an intrinsic feature of any multicenter problem, and not a defect of the methodology described.

**Acknowledgments**
The authors are grateful to J. Phys. B Referee for useful comments concerning Appendix A. This work was supported by the Uzbek Foundation Award OT-Ф2-46 (ASB) and U.S. DOE, Basic Energy Sciences, Office of Energy Research (AZM and STM).

**Appendix A**

The spherical *s*- and *p*- waves propagating from the point $\mathbf{R}_j$ can be represented as a result of the differential operators $\hat{B}_{lm}^{(j)}$ acting on the Green's function of a free particle

$$G_k^+(\mathbf{r}, \mathbf{R}_j) = \frac{e^{ik|\mathbf{r}-\mathbf{R}_j|}}{2\pi|\mathbf{r}-\mathbf{R}_j|}. \qquad (A1)$$

For the *s*-waves this operator is simply $\hat{B}_{00}^{(j)} = \sqrt{\pi}$.
For the *p*-waves these operators have the following forms

$$\hat{B}^{(j)}_{1-1} = \sqrt{\frac{3\pi}{2}} \frac{1}{k} \left( \frac{\partial}{\partial R^{(j)}_x} - i\frac{\partial}{\partial R^{(j)}_y} \right);$$

$$\hat{B}^{(j)}_{10} = \sqrt{3\pi} \frac{\partial}{k\partial R^{(j)}_z};$$

$$\hat{B}^{(j)}_{11} = -\hat{B}^{(j)*}_{1-1}, \tag{A2}$$

where $R^{(j)}_x, R^{(j)}_y, R^{(j)}_z$ are the Cartesian components of the vector $\mathbf{R}_j$.

For the *d*-waves these operators have the following forms
(below $\boldsymbol{\rho} = \mathbf{r} - \mathbf{R}_j$; $\rho_x = x - R^{(j)}_x$ and the same for *y* and *z*)

$$\hat{B}^{(j)}_{2-2} = \frac{1}{4}\sqrt{30\pi}\frac{\rho^2}{k^2}\left[\hat{F}_x - \hat{F}_y - 2i\hat{F}_{xy}\right];$$

$$\hat{B}^{(j)}_{22} = \hat{B}^{(j)*}_{2-2};$$

$$\hat{B}^{(j)}_{2-1} = \frac{1}{4}\sqrt{120\pi}\frac{\rho^2}{k^2}\left[\hat{F}_{xz} - i\hat{F}_{yz}\right];$$

$$\hat{B}^{(j)}_{21} = -\hat{B}^{(j)*}_{2-1};$$

$$\hat{B}^{(j)}_{20} = \frac{1}{4}\sqrt{20\pi}\frac{\rho^2}{k^2}\left[2\hat{F}_z - \hat{F}_x - \hat{F}_y\right] \tag{A3}$$

where

$$\hat{F}_x = \left[\left(\frac{1}{\rho}\frac{\partial}{\partial R^{(j)}_x}\right)^2 - \frac{1}{\rho^4}(1-ik\rho)\left(\frac{\rho^2_x}{\rho^2} - 1\right)\right], \text{ and the same for } y \text{ and } z;$$

$$\hat{F}_{xy} = \left[\left(\frac{1}{\rho}\frac{\partial}{\partial R^{(j)}_x}\right)\left(\frac{1}{\rho}\frac{\partial}{\partial R^{(j)}_y}\right) - \frac{1}{\rho^4}(1-ik\rho)\left(\frac{\rho_x}{\rho}\right)\left(\frac{\rho_y}{\rho}\right)\right], \text{ and the same for pairs } xz \text{ and } yz.$$

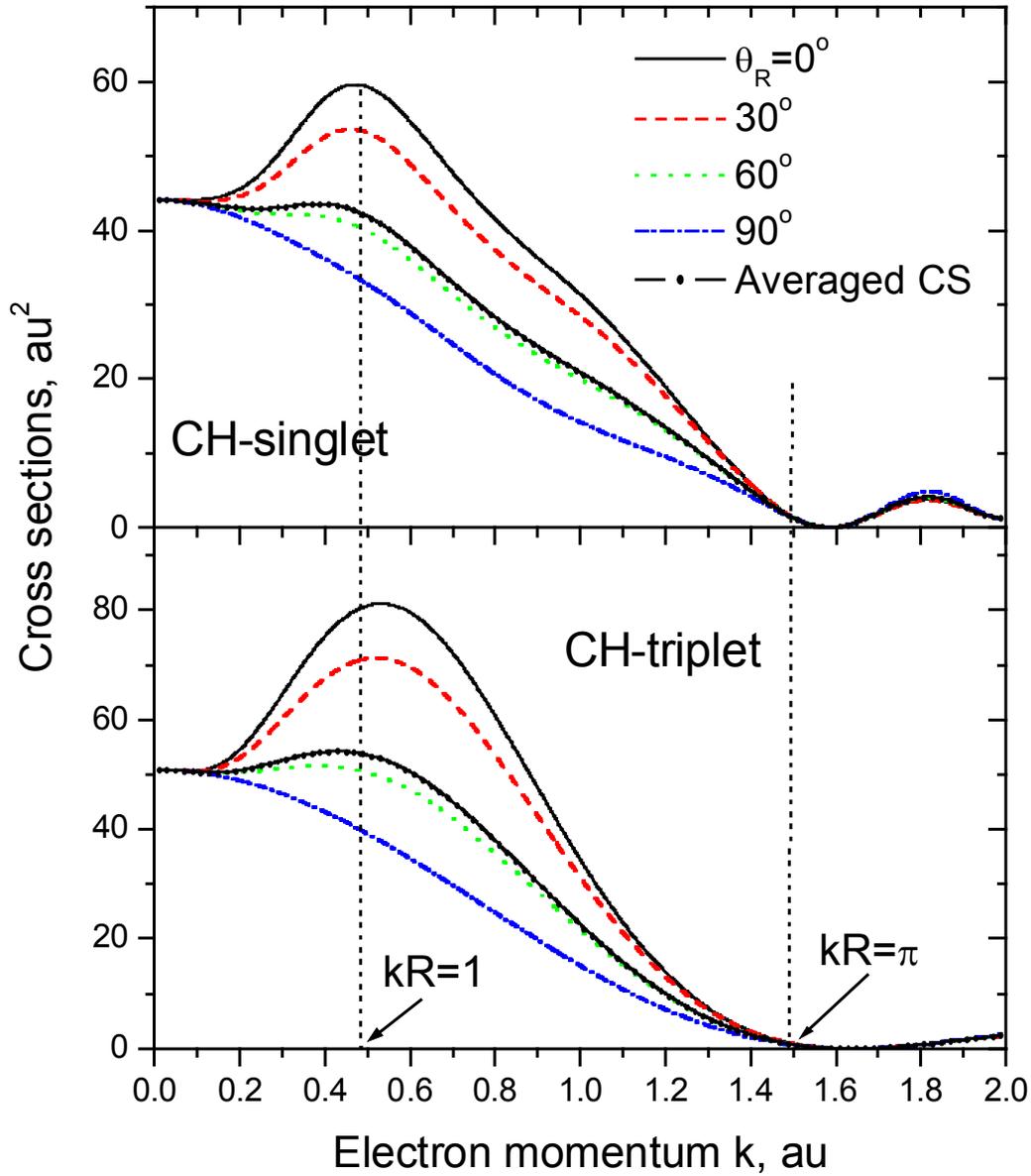

Fig. 1. Total $\sigma_{tot}(\mathbf{k},\mathbf{R})$ and averaged $\bar{\sigma}(k)$ (Averaged CS) elastic electron scattering cross sections (in atomic units au$^2$) of CH molecules as a function of electron momentum $k$; $\vartheta_R$ is the angle between the molecular axis vector $\mathbf{R}$ and $\mathbf{k}$. Results for averaged cross sections $\bar{\sigma}(k)$ almost coincide with those of $\sigma_{tot}(\mathbf{k},\mathbf{R})$ for $\vartheta_R = 60^o$.

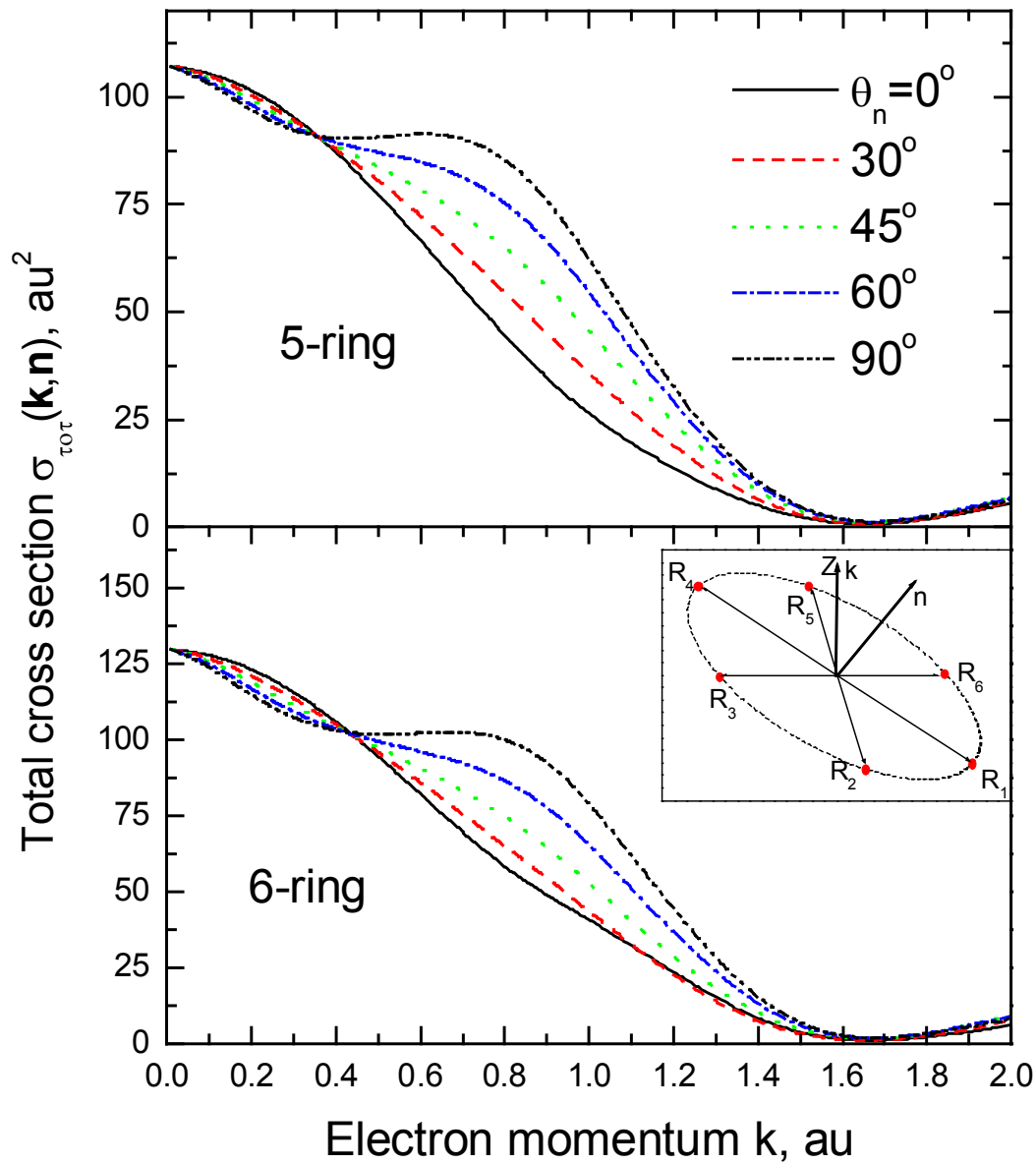

Fig. 2. Total elastic electron scattering cross section $\sigma_{tot}(\mathbf{k}, \{\mathbf{R}_j\})$ for different positions of 5- and 6-carbon rings relative to incident electron beam $\mathbf{k}$ as a function of electron momentum $k$. The vector $\mathbf{n}$ (in the insert in the lower panel) is the normal to the plane where the 6-ring (or 5-ring) carbon is located; $\vartheta_n$ is the angle between the vectors $\mathbf{n}$ and $\mathbf{k}$ (or z-axis). In the rotation of the ring plane the vectors $\mathbf{k}$, $\mathbf{n}$ and $\mathbf{R}_1$ are located in the same plane.

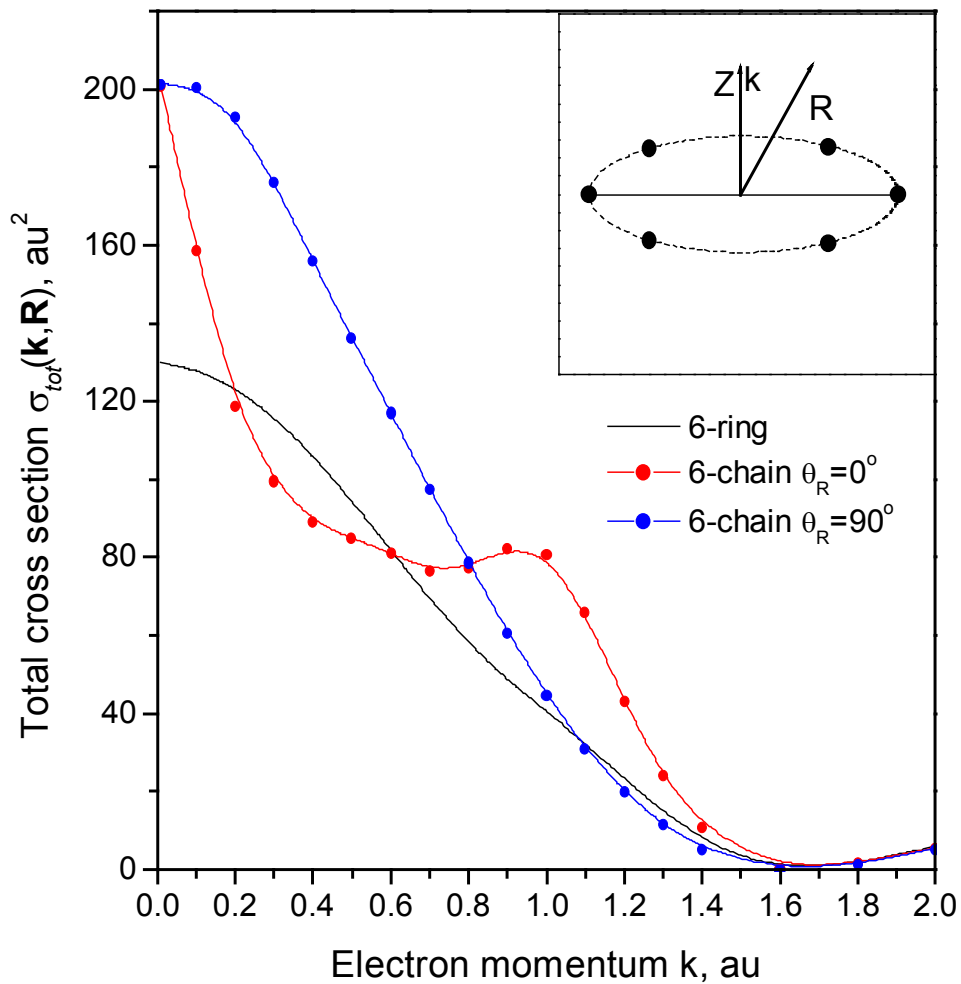

Fig. 3. Total elastic electron scattering cross section $\sigma_{tot}(\mathbf{k},\{\mathbf{R}_j\})$ for $C_6$ chain as a function of electron momentum $k$ for two different positions of chain axis **j** relative to the incident electron beam **k**. The vector **R** (insert in the upper panel) is the chain axis; $\vartheta_R$ is the angle between the vectors **k** and **R**. For comparison, the solid curve is the total elastic electron scattering cross section by the 6-ring carbon molecule located in the plane perpendicular to the vector **k**.

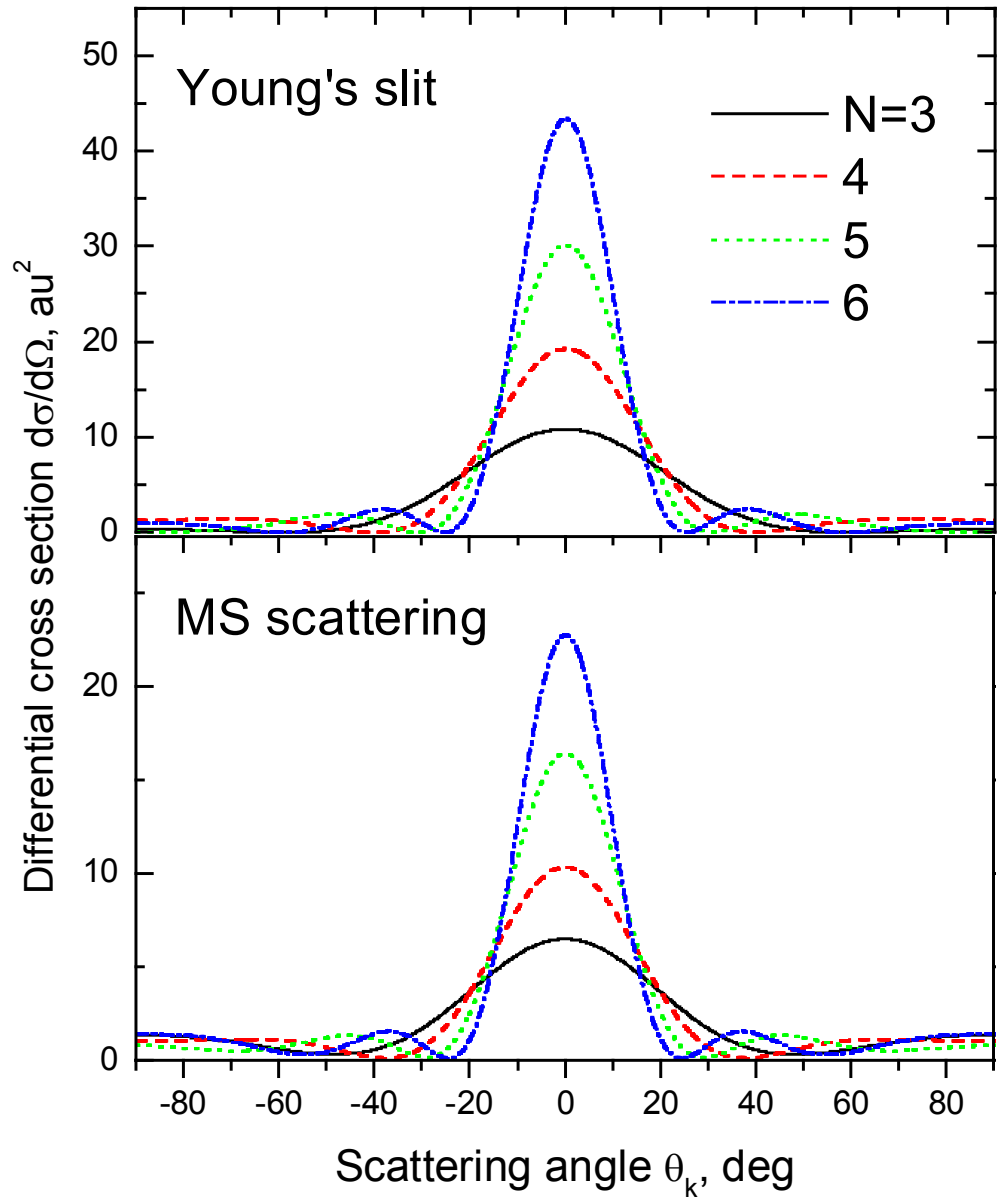

Fig. 4. Differential cross sections $d\sigma/d\Omega_{k'}$ for electron scattering by the linear chains of carbon atoms; $N$ is the number of C atoms in the chain and the electron momentum $k$=0.9 au. The axis of the atomic chain **j** is perpendicular to the incident electron beam, **k**. The vectors **k, k'** and **j** are located in the plane of the figure. The upper panel represents the single wave scattering approximation; the lower one is the multiple-scattering approximation.